\documentclass[12pt]{article}
\usepackage{enumerate}
\usepackage{natbib}
\setcitestyle{authoryear,open={(},close={)}}
\usepackage{array, booktabs, multirow}
\newcolumntype{C}[1]{>{\centering\arraybackslash}p{#1}}
\usepackage{graphicx,psfrag,epsf}
\usepackage{bm}
\usepackage{amsmath}
\usepackage{amssymb}
\usepackage{bbm}
\usepackage{float}
\usepackage[utf8]{inputenc}
\usepackage{tabulary}
\usepackage{setspace}
\usepackage{xcolor}
\usepackage{subcaption}
\usepackage{chngpage}
\usepackage{textcomp}
\usepackage{gensymb}
\usepackage{url}
\usepackage{natbib}
\usepackage{algorithm}
\usepackage{algpseudocode}
\usepackage{authblk}
\usepackage{blindtext}
\usepackage{longtable}
\usepackage{comment}

\makeatletter
\renewcommand\@biblabel[1]{#1.}
\makeatother

\usepackage{mathtools}
\newcommand{\blind}{0}

\newcommand{\T}{\top}
\newcommand{\given}{\, | \,}

\newcommand{\pari}{\hspace{\parindent}}

\begin{document}

\def\spacingset#1{\renewcommand{\baselinestretch}%
{#1}\small\normalsize} \spacingset{1}


\if0\blind
{
\title{\bf A Unifying Bayesian Approach for Sample Size Determination Using Design and Analysis Priors 
}
\author{Jane Pan\\
Department of Biostatistics, University of California, Los Angeles (UCLA)\\
and \\
Sudipto Banerjee \\
Department of Biostatistics, University of California, Los Angeles (UCLA)}
\maketitle
} \fi

\if1\blind
{
\bigskip
\bigskip
\bigskip
\begin{center}
{\LARGE\bf Title}
\end{center}
\medskip
} \fi

\bigskip
\begin{abstract}
  Power and sample size analysis comprises a critical component of clinical trial study design. There is an extensive collection of methods addressing this problem from diverse perspectives. The Bayesian paradigm, in particular, has attracted noticeable attention and includes different perspectives for sample size determination.
  Building upon a cost-effectiveness analysis undertaken by \cite{ohagan} with different priors in the design and analysis stage, we develop a general Bayesian framework for simulation-based sample size determination that can be easily implemented on modest computing architectures. We further qualify the need for different priors for the design and analysis stage. We work primarily in the context of conjugate Bayesian linear regression models, where we consider the situation with known and unknown variances. Throughout, we draw parallels with frequentist solutions, which arise as special cases, and alternate Bayesian approaches with an emphasis on how the numerical results from existing methods arise as special cases in our framework.
\end{abstract}

\noindent%
\vfill

\spacingset{1.45}


\section{Introduction}\label{Intro}

\pari A crucial problem in experimental design is that of determining the
sample size for a proposed study.
There is, by now, a substantial literature in classical and Bayesian settings. Classical sample size calculations have been treated in depth
in texts such as \cite{thiemann},
\cite{cohen}, and \cite{desu}, while extensions to linear and generalized linear models
have been addressed in \cite{self}, \cite{ohara}, \cite{muller}
and \cite{liu}. The Bayesian setting has also allocated a decent amount of attention towards sample size determination. A widely referenced issue of
\emph{The Statistician} (vol. 46, issue 2, 1997) includes a number of articles
from different perspectives \citep[see, e.g., the articles by][]{lindley, pham, cj, joseph}. Within the Bayesian setting itself, there have been efforts to distinguish between a formal
utility approach  \citep{raiffa, berger, lindley, parm} and approaches that
attempt to determine sample size based upon some criterion of analysis or
model performance \citep{rahme, wang, ohagan}.
Other proposed solutions adopt a more tailored approach. For example, \cite{Ibrahim} specifically target Bayesian meta-experimental design using survival regression models; \cite{reyes} propose a framework based on Bayesian  average errors capable of simulatenously controlling for Type I and Type II errors, while \cite{joseph} rely on lengths of posterior credible intervals to gauge their sample size estimates.

The aforementioned literature presents the problem in a variety of applications including, but not limited to, clinical trials. Bayesian treatments specific to clinical trials can be found in \cite{spiegelhalter}, \cite{parmigiani}, \cite{berry}, \cite{berry2010}, \cite{lee_zelen}, and \cite{lee}. Regardless of the specific approach, all of the cited articles above address the sample size problem based on some well-defined objective that is desired in the analysis stage. The design of the study, therefore, should consider that the analysis objective is met with a certain probability. The framework we develop here is built upon this simple idea. A clear analysis objective and proper sampling execution are all that is needed to provide us with the necessary sample size.

\subsection{Classical Power and Sample Size}\label{frequentist}

\begin{figure}[htb]
\centering
\includegraphics[width=0.45\textwidth]{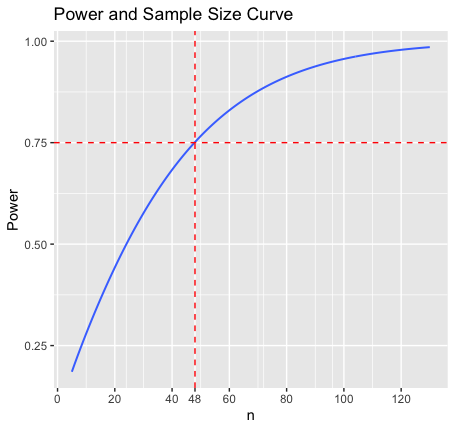}
\caption{Power curve example. Dotted lines indicate how a sample size of 34 is required to achieve
  a power of 0.75.}
\label{fig:powercurve}
\end{figure}

The classical power and sample size analysis problem in the frequentist setting is widely applied in diverse settings. For example, we can formulate a hypothesis concerning the population mean. A one-sided hypothesis test for the null $\displaystyle H_0: \theta = \theta_0$ against the alternative $H_a: \theta > \theta_0$, where $\theta$ is the population mean, is decided on the location of $\theta_0$ with respect to the distribition of the sample mean $\bar{y}$. Assuming a known value for the population variance $\sigma^2$ and that the sample mean's distribition is (approximately) Gaussian, we reject $H_0$ if $\bar{y} > \theta_0 + \frac{\sigma}
{\sqrt{n}}Z_{1-\alpha}$, where $\alpha$ is the specified Type-I error and $Z_{1-\alpha}$ is the corresponding quantile of the Gaussian distribution. The statistical ``power'' of the test is $1-\beta$, where $\beta = P(\mbox{Rej } H_0\given H_a)$ is the Type-II error. Straightforward algebra yields the ubiquitous sample size
formula, $\displaystyle n = (Z_{\alpha} + Z_{\beta})^2\left(\frac{\sigma}{\Delta}\right)^2$, derived from
the power:
\begin{equation} \label{eq:freq_power}
1-\beta = P\left(\bar{y} > \theta_0 +  \frac{\sigma}{\sqrt{n}}Z_{1-\alpha}
  \given \theta = \theta_1\right) = \Phi\left(\sqrt{n}\frac{\Delta}{\sigma} + Z_{\alpha}\right),
\end{equation}
where $\Delta = \theta_1 - \theta_0$ is the critical difference. Given any fixed value of $\Delta/\sigma$, the power curve is a function of sample size and can be plotted as in Figure~\ref{fig:powercurve}. The sample size required to achieve a specified power can then be read from the power curve.

More generally, power curves, such as the one shown in Figure~\ref{fig:powercurve}, may not be available in closed form but can be simulated for different sample sizes. They quantify our degree of assurance regarding our ability to meet our analysis objective (rejecting the null) for different sample sizes. Formulating sample size determination as a decision problem so that power is an increasing function of sample size, offers the investigator a visual aid in helping deduce the
minimum sample size needed to achieve a desired power.

\subsection{Bayesian Assurance}\label{bayesian}


From a Bayesian standpoint, there is no need to condition on a fixed alternative. Instead, we determine the tenability of a hypothesis given the data we observe. A joint probability model is constructed for the parameters and the data using a prior distribition for the parameters and the likelihood function of the data conditional on the parameters. Inference proceeds from the posterior distribition of the parameters given the data.

In the design stage we have not observed the data. Therefore, we will need to formulate a data generating mechanism and, subsequently, consider the posterior distribution given the realized data to evaluate the tenability of a hypothesis regarding our parameter. We then use the probability law associated with the data generating mechanism to assign a degree of assurance of our analysis objective. As a sufficiently simple example, consider a situation where we seek to evaluate the tenability of
$H: \theta > \theta_0$ given data from a Gaussian population with mean $\theta$ and a known variance $\sigma^2$. Assuming the prior $\theta \sim N\left(\theta_1, \frac{\sigma^2}{n_0}\right)$,
where $n_0$ reflects the precision of the prior relative to the data, and the likelihood of the sample mean $\displaystyle \bar{y} \sim N\left(\theta, \frac{\sigma^2}{n} \right)$, the posterior distribution of $\theta$ can then be obtained by multiplying the prior and the likelihood as
\begin{equation}\label{eq: simple_posterior}
N\left(\theta {\left | \theta_1, \frac{\sigma^2}{n_0} \right.}\right) \times N\left(\bar{y} {\left | \theta, \frac{\sigma^2}{n} \right.}\right)
\propto N\left(\theta {\left | \frac{n_0}{n + n_0}\theta_1 + \frac{n}{n + n_0}\bar{y}, \frac{\sigma^2}{n + n_0}\right.}\right)\; .
\end{equation}
Our analysis objective is to ascertain whether $P(\theta > \theta_0 \given \bar{y}) > 1-\alpha$, where $\alpha$ is a fixed number. The posterior distribution in (\ref{eq: simple_posterior}) gives us $P(\theta > \theta_0 \given \bar{y})$ and we define
\begin{equation}
\label{eq:probofprob}
\delta = P_{\bar{y}}\left\{\bar{y}: P(\theta > \theta_0 \given \bar{y}) > 1 - \alpha\right\}
\end{equation}
as a Bayesian counterpart of statistical power, which we refer to as \emph{Bayesian assurance}. The inner probability defines our analysis objective, while the outer probability defines our chances of meeting the analysis objective under the given data generating mechanism.

\subsection{Manuscript Overview}\label{aims}

\pari Our current manuscript intends to explore Bayesian assurance and subsequent sample size calculations as described through (\ref{eq: simple_posterior})~and~(\ref{eq:probofprob}) in the general context of conjugate Bayesian linear regression. Of particular emphasis will be the data generating mechanism and providing motivation behind quantifying separate prior beliefs at the design and analysis stage of clinical trials  \citep{ohagan}. 
The balance of this manuscript proceeds as follows. Section~\ref{BayesianSampleSizeSec} presents the Bayesian sample size determination problem embedded within a conjugate Bayesian linear model. This offers us analytic tractability through which we motivate the need for different prior distributions for designing and analyzing a study.
Section~\ref{subsec: twocases} considers the cases of known and unknown variances and offers corresponding algorithms.
Section~\ref{two_stage_apps_section} casts the cost-effectiveness problem explored by \cite{ohagan} in our framework, which we revisit using known and unknown variances, and we also compare with other Bayesian alternatives for sample size calculations including for inference using proportions. We close the paper in Section~\ref{sec: conclusions} with additional points of discussion and future aims.

\section{Bayesian Assurance and Sample Size Determination}\label{BayesianSampleSizeSec}
%
\subsection{Conjugate Bayesian Linear Regression}\label{sec: bayesian_linear_regression}
Consider a proposed study where a certain number, say $n$, of observations, which we denote by
$y_n = (y_1,y_2 \ldots,y_n)^{\top}$, are to be collected in the presence of $p$ controlled explanatory variables, say $x_1, x_2,\ldots,x_p$, that will be known to the investigator for any unit $i$ at the design stage. Consider the usual normal linear regression setting
such that $y_n = X_n\beta + \epsilon_n$, where $X_n$ is $n\times p$ with $i$-th row corresponding to $x_i^{\top}$ and $\epsilon_n \sim N(0,\sigma^2V_n)$, where $V_n$ is a known $n\times n$ correlation matrix. Both $X_n$ and $V_n$ are assumed known for each sample size $n$ through design and modeling considerations. A conjugate Bayesian linear regression model specifies the joint distribution of the parameters $\{\beta,\sigma^2\}$ and the data as
\begin{equation}\label{eq: bayesian_conj_regression}
  IG(\sigma^2\given a_{\sigma}, b_{\sigma}) \times
  (\beta \given \mu_{\beta},\sigma^2 V_{\beta})
  \times N(y_n\given X_n\beta,\sigma^2 V_n)                                            \end{equation}                                                                           Inference proceeds from the posterior distribution derived from (\ref{eq: bayesian_conj_regression}),
\begin{align}\label{eq: Conjugate_Bayesian_LM_Posterior}                                     p(\beta, \sigma^2 \given y_n) &= \underbrace{IG(\sigma^2\given
    a^*_{\sigma}, b^*_{\sigma})}_{p(\sigma^2\given y_n)} \times
    \underbrace{N(\beta\given M_nm_n, \sigma^2 M_n)}_{p(\beta\given \sigma^2, y_n)} \;,  \end{align}                                                                              where $a^*_{\sigma} = a_{\sigma}+n/2$, $b^*_{\sigma} = b_{\sigma} +
(1/2)\left\{\mu_{\beta}^{\top}V_{\beta}^{-1}\mu_{\beta} +
y_n^{\top}V_n^{-1}y_n - m_n^{\top}M_nm_n\right\}$, $M_n^{-1} =
V_{\beta}^{-1} + X_n^{\top}V_n^{-1}X_n$ and $m_n =
V_{\beta}^{-1}\mu_{\beta} + X_n^{\top}V_n^{-1}y_n$.
Sampling from the joint posterior distribution of
$\{\beta,\sigma^2\}$ is achieved by first sampling
$\sigma^2 \sim IG(a^*_{\sigma}, b^*_{\sigma})$ and
then sampling $\beta \sim N(M_nm_n, \sigma^2 M_n)$
for each sampled $\sigma^2$. This yields marginal
posterior samples from $p(\beta\given y_n)$, which
is a non-central multivariate $t$ distribution, though
we do not need to work with its complicated density
function. See \cite{gelman}
for further details on the conjugate Bayesian linear
regression model and sampling from its posterior.

We wish to decide whether our realized data will favor
$H: u^{\top}\beta > 0$, where $u$ is a $p\times 1$ vector of fixed contrasts. In practice, a decision on the tenability of $H$ is often based on the $100(1-\alpha)\%$ posterior credible interval,
\begin{equation}                                                                            \left(u^{\top}M_nm_n - Z_{1-\alpha/2}\sigma\sqrt{u^{\top}M_nu},\;
   u^{\top}M_nm_n + Z_{1-\alpha/2}\sigma\sqrt{u^{\top}M_nu}\right)\;,                   \end{equation}                                                                           obtained from the conditional posterior predictive distribution
$p(\beta\given \sigma, y_n)$. The data would favor $H$ if $y_n$ belongs
to the following set
\[
S_{\alpha}(n; y_n,\sigma, \mu_{\beta}, V_{\beta}, V_n) = \left\{y_n :
u^{\top}M_mm_n > Z_{1-\alpha/2}\sigma\sqrt{u^{\top}M_nu}
\right\}\;.
\]
This is equivalent to $0$ being below the two-sided $100(1-\alpha)\%$ credible interval for $u^{\top}\beta$. Practical Bayesian designs will seek to assure the investigator that the above criterion will be achieved with a sufficiently high probability through the \emph{Bayesian assurance},                                              \begin{multline}\label{eq: bayesian_assurance}
\delta(n; \sigma, u, \mu_{\beta}, V_{\beta}, V_n) = P_{y_n}(S_{\alpha}(n; y_n, \sigma, \mu_{\beta}, V_{\beta}, V_n)) \\ = P_{y_n}\left\{y_n : u^{\top}M_mm_n >                  Z_{1-\alpha/2}\sigma\sqrt{u^{\top}M_nu} \right\}\;,
\end{multline}
which generalizes the definition in (\ref{eq:probofprob}). Given the assumptions on the model, the fixed values of the parameters $\{\mu_{\beta}, V_{\beta},\sigma, V_n\}$ and the fixed vector $u$ that determines the hypothesis being tested, the Bayesian assurance function evaluates the probability of rejecting the null hypothesis under the marginal probability distribution of the realized data corresponding to any given sample size $n$. Choice of sample size will be determined by the smallest value of $n$ that will ensure $\delta(n; \sigma, u, \mu_{\beta}, V_{\beta}, V_n) > \gamma$, where $\gamma$ is a specified number.

\subsection{Limitations for a single prior}\label{sec:limitations}

Let us consider the special case when $X_n = 1_n$ so that $\beta$ is a scalar with prior distribution $\beta \sim N(\beta_1, \sigma^2/n_0)$, where $n_0 > 0$ is a fixed precision parameter (sometimes referred to as ``prior sample size''), $V_n = I_n$ and $H : \beta > \beta_0$. We decide in favor of $H$ if the data lies in
\begin{equation*}
S_{\alpha}(n; y, \sigma, \beta_0, \beta_1, n_0) 
 = \left\{\bar{y} : \bar{y} > \beta_0 - \frac{n_0}{n}(\beta_1 - \beta_0) - \sqrt{\left(1 + \frac{n_0}{n}\right)}\frac{\sigma}{\sqrt{n}}Z_{\alpha} \right\}\;,
\end{equation*}
where the expression on the right reveals a convenient condition in terms of the sample mean. As $n_0\to 0$, i.e., the prior becomes vague, $S_{\alpha}(n; y, \sigma, \beta_0, \beta_1, n_0)$ collapses to the critical region in classical inference for testing $H_0 : \beta = \beta_0$ against $H_a : \beta = \beta_1$. The Bayesian assurance function is
\begin{equation}\label{eq: bayesian_assurance_simple_null}                                                                                                                                                                                                                                                        \delta(n; \sigma, \Delta,n_0) = \Phi\left(\sqrt{n_0}\left[\sqrt{1+\frac{n_0}{n}}\left(\frac{\Delta}{\sigma}\right) + Z_{\alpha}\sqrt{\frac{1}{n}} \right]\right)\;,                                                                                                                                                                                                                                                        \end{equation}                                                                                                                                                                                                                                                          where $\Delta = \beta_1 - \beta_0$. Given $n_0$, we will compute the sample size needed to detect a critical difference of $\Delta$ with probability $1-\beta$  as
$n = \arg\min\{n: \delta(\Delta,n) \geq 1-\beta\}$.                                                                                                                                                                                                                                                          However, the limiting properties of the function in (\ref{eq: bayesian_assurance_simple_null}) are not without problems. When the prior is vague, i.e., $n_0 \to 0$, then
\[                                                                                                                                                                                                                                                            \lim_{n_0\to 0} \delta(\Delta,n) = \Phi\left(0\right) = 0.5\;,                                                                                                                                                                                                                                                            \]
while in the case when the prior is precise, i.e., $n_0 \to \infty$ we obtain
\begin{equation}\label{eq: paradox}                                                                                                                                                                                                                                                        \lim_{n_0\to \infty} \delta(\Delta,n) = \left\{                                                                                                                                                                                                                                                            \begin{array}{ll}                                                                                                                                                                                                                                                        1 & \mbox{if $\Delta > 0$}\\                                                                                                                                                                                                                                                        0 & \mbox{if $\Delta \leq 0$} \\                                                                                                                                                                                                                                                        \end{array} \right. .                                                                                                                                                                                                                                                        \end{equation}
This is undesirable. Vague priors are customary in Bayesian analysis, but they propagate enough uncertainty that the marginal distribution of the data under the given model will force the assurance to be lower than 0.5. In other words, regardless of how large a sample size we have, we cannot assure the investigator with probability greater than 50\% that the null hypothesis will be rejected. At the other extreme, where the prior is fully precise, it fully dominates the data (or the likelihood) and there is no information from the data that is used in the decision. Therefore, the assurance is a function of the prior only and we will always or never reject the null hypothesis depending upon whether $\Delta > 0$ or $\Delta < 0$.

In order to resolve this issue, we work with two different sets of priors, one at the \emph{design} stage and another at the \emph{analysis stage}. Building upon O'Hagan and Stevens (2001), we elucidate with the Bayesian linear regression model in the next section and offer a simulation-based framework for computing the Bayesian assurance curves.

\subsection{Bayesian Assurance Using Design and Analysis Priors}\label{subsec: twocases}
\pari We consider two scenarios that are driven by the amount of information given in a study. We develop the corresponding algorithms  based on these assumptions. The first case assumes that the population variance $\sigma^2$ is known
and the second case assumes $\sigma^2$ is unknown, prompting us to consider additional prior distributrions for $\sigma^2$ in the design and analysis stage. Our context remains testing the tenability of $H: u^{\T}\beta > C$ given realized data from a study, where $C$ is a known constant.

\subsubsection{Known Variance $\sigma^2$}\label{subsec: ohagan_stevens_linear_regression_known_sigma}

If $\sigma^2$ is known and fixed, then the posterior distribution of $\beta$ is
$\displaystyle p(\beta \given \sigma^2, y_n) = N(\beta | M_nm_n, \sigma^2M_n)$ as shown
in Equation~\ref{eq: Conjugate_Bayesian_LM_Posterior}. Hence, standardization leads to
\begin{equation} \label{eq1}                                                                    \left. \frac{u^{\top}\beta - u^{\top}M_nm_n}{\sigma \sqrt{u^{\top}M_nu}} \right| \sigma^2, y_n \sim N(0,1)\;.
\end{equation}
To evaluate the credibility of $H: u^{\top}\beta > C$, where $u$ denotes
a known $p \times 1$ vector and $C$ is a known constant,
we decide in favor of $H$ if the observed data belongs in the region:
\begin{align*}
    A_{\alpha}(u, \beta, C) &= \left\{y_n: P\left(u^{\top}\beta
    \leq C | y_n\right)  < \alpha\right\} = \left\{y_n: \Phi
    \left(\frac{C - u^{\top}M_nm_n}{\sigma \sqrt{u^{\top}M_nu}}\right)
    < \alpha \right\}.
\end{align*}
Given the data $y_n$ and the fixed parameters in the analysis priors, we can evaluate $M_n$ and $m_n$ and
hence, for any given $\sigma$, $C$ and $\alpha$, ascertain if we have credibility for $H$ or not.

In the {design objective} we need to ask ourselves ``What sample size is needed to assure us that the analysis objective is met $100\gamma\%$ of the time?'' Therefore, we seek $n$ so that 
\begin{align}\label{eq: bayesian_assurance_2}
    \delta(n) &= P_{y_n}(A_{\alpha}(u, \beta, C)) = P_{y_n}\left\{y_n:
    \Phi \left(\frac{C - u^{\top}M_nm_n}{\sigma \sqrt{u^{\top}M_nu}}
    \right) < \alpha \right\} \geq \gamma\;,
\end{align}
where $\delta(n)$ is the Bayesian assurance. In order to
evaluate (\ref{eq: bayesian_assurance_2}), we will need the
marginal distribution of $y_n$. In light of the paradox in
(\ref{eq: paradox}), our belief about the population from
which our sample will be taken is quantified using the design
priors. Therefore, the ``marginal'' distribution of $y_n$ under
the design prior will be derived from
\begin{align}\label{eq: ohagan_stevens_linear_regression_design_priors}
  y_n &=  X_n\beta + e_n; \quad e_n \sim N(0, \sigma^2 V_n)\;;\quad \beta = \mu_{\beta}^{(d)} + \omega; \quad \omega \sim N(0, \sigma^2 V_{\beta}^{(d)})\;,
\end{align}
where $\beta\sim N(\mu_{\beta}^{(d)},\sigma^2 V_{\beta}^{(d)})$ is the design prior on $\beta$. Substituting the equation for $\beta$ into the equation for $y_n$ in (\ref{eq: ohagan_stevens_linear_regression_design_priors}) gives $y_n = X\mu_{\beta}^{(d)} + (X\omega + e_n)$ and, hence, $y_n \sim N\left(X\mu_{\beta}^{(d)}, \sigma^2V_n^{*}\right)$, where $V_n^{\ast} = \left(XV_{\beta}^{(d)}X^{\top} + V_n\right)$. We now have a simulation strategy to estimate our Bayesian assurance. We fix sample size $n$ and generate a sequence of $J$ data sets $y_n^{(1)}, y_n^{(2)},\ldots, y_n^{(J)}$, each of size $n$ from $N\left(X\mu_{\beta}^{(d)}, \sigma^2V_n^{*}\right)$. Then, a Monte Carlo estimate of the Bayesian assurance is
\begin{equation}\label{eq: bayesian_assurance_estimate}
    \hat{\delta}(n) = \frac{1}{J}\sum_{j=1}^J \mathbb{I}\left(\left\{y_n^{(j)}:
    \Phi \left(\frac{C - u^{\top}M_n^{(j)}m_n^{(j)}}{\sigma
    \sqrt{u^{\top}M_n^{(j)}u}}\right) < \alpha \right\}\right)\;,
\end{equation}
where $\mathbb{I}(\cdot)$ is the indicator function of the
event in its argument, $M_n^{(j)}$ and $m_n^{(j)}$ are the
values of $M_n$ and $m_n$ computed from $y_n^{(j)}$.
We repeat the steps needed to compute (\ref{eq: bayesian_assurance_estimate})
for different values of $n$ and obtain a plot of $\delta(n)$ against $n$.
Our desired sample size is the smallest $n$ for which $\hat{\delta}(n)
\geq \gamma$, where we seek assurance of a $100\gamma\%$ chance of
deciding in favor of $H$.

A special case of the model can be considered
where $X_n = 1_n$ is an $n\times 1$ vector of ones,
$\beta$ is a scalar, $V_n = I_n$ and we wish to evaluate the credibility of
$H : \beta > \beta_0$. We assume $\beta \sim N(\beta_1, \sigma^2/n_a)$
in the analysis stage and $\beta \sim N(\beta_1, \sigma^2/n_d)$ in the
design stage, where $\beta_1 > \beta_0$. The data will favor $H$ if
the sample mean lies in $A_{\alpha}(\beta_0,\beta_1)$, where
\[
A_{\alpha}(\beta_0,\beta_1) = \left\{\bar{y}:  \bar{y} >  \beta_0 -
\frac{n_a}{n}(\beta_1-\beta_0) - \sqrt{\left(1+\frac{n_a}{n}
\right)}\frac{\sigma}{\sqrt{n}} Z_{\alpha} \right\}\;.
\]
Using the design prior, we obtain the marginal distribution $\displaystyle \bar{y}\sim N\left(\beta_1, \left(\frac{1}{n} + \frac{1}{n_d}\right)\sigma^2\right)$. We use this distribution to calculate $\delta(n) = P_{\bar{y}}\{\bar{y}: P(\theta < \theta_0 \given\bar{y}) < \alpha\}$, which produces a closed-form expression for Bayesian assurance:
\begin{equation}\label{eq: bayesian_assurance_3}
    \delta(\Delta, n, n_a, n_d) = \Phi\left(\sqrt{\frac{nn_d}{n + n_d}}
    \left[\frac{n+n_a}{n}\frac{\Delta}{\sigma} + Z_{\alpha}
    \frac{\sqrt{n + n_a}}{n} \right]\right)\;,
\end{equation}
where $\Delta = \beta_1 - \beta_0$. As $n_d\to \infty$ and $n_a\to 0$,
we obtain that
\[
\lim_{n_a\to 0,n_d\to\infty} = \Phi\left(\sqrt{n}\frac{\Delta}{\sigma} +
Z_{\alpha}\right)\;,
\]
which is precisely the frequentist power curve. Therefore,
the frequentist sample size emerges as a special case of the
Bayesian sample size when the design prior becomes perfectly
precise and the analysis prior becomes perfectly uninformative. Algorithm~\ref{alg:known_var} presents a pseudocode to compute Bayesian assurance.

\begin{algorithm}
\caption{Bayesian assurance algorithm for known variance}
\label{alg:known_var}
\begin{algorithmic}[1]
\Procedure{bayes sim}{$n$, $u$, $C$, $X$, $V_n$, $V_{\beta}^{(d)}$,
  $V_{\beta}^{-1(a)}$, $\mu_{\beta}^{(d)}$, $\mu_{\beta}^{(a)}$, $\sigma^2$, $\alpha$}
\State{count = $0$} \Comment{keeps track of iterations
  satisfying the analysis objective}\\
\For{$i$ in range $1$ : max number of iterations}
\State{\textbf{Design Stage Starts}}
\State{$y \gets$ Vector of $n$ values each generated
  from N$(X\mu_{\beta}^{(d)}, \sigma^2(X V_{\beta}^{(d)}X^{\top} + V_n))$}
\State{\textbf{Design Stage Ends}}\\
\State{\textbf{Analysis Stage Starts}} \Comment{Computes parameters
  of the $\beta$ posterior:}
\State{$M \gets (V_\beta^{-1(a)} + X^{\top} V_n^{-1} X)^{-1}$}
\State{$m \gets V_\beta^{-1(a)}\mu_\beta^{(a)} + X^{\top}V_n^{-1}y$}
\If{$\frac{C - u^{\top}Mm}{\sigma \sqrt{u^{\top}Mu}} < Z_\alpha$}
\State{$Z_i \gets 1$}
\Else
\If{$\frac{C - u^{\top}Mm}{\sigma \sqrt{u^{\top}Mu}} \geq Z_\alpha$}
\State{$Z_i \gets 0$}
\EndIf
\EndIf\\
\State{count $\gets$ count + $Z_i$}
\State{\textbf{Analysis Stage Ends}}
\EndFor\\
\State{\textbf{assurance} $\gets$ count / max number of iterations}\\
\Return{assurance}\\
\EndProcedure
\end{algorithmic}
\end{algorithm}

\subsubsection{Unknown Variance $\sigma^2$}\label{subsec:unknownvar}
When $\sigma^2$ is unknown, the posterior distribution of interest is $p(\beta, \sigma^2 \given y_n)$ as opposed to the original $p(\beta \given \sigma^2, y_n)$ delineated in the known variance case.
Since $\sigma^2$ is no longer fixed, it becomes challenging to define a closed form condition that is capable of evaluating the credibility of $H: u^{\top}\beta > C$. Hence, we do not obtain a condition similar to \eqref{eq1}. However, our region of interest corresponding to our analysis objective still remains as $A_{\alpha}(u, \beta, C) = \left\{y_n: P\left(u^{\top}\beta \leq C \given y_n\right)  < \alpha\right\}$ when deciding whether or not we are in favor of $H$. To implement this in a simulation setting, we rely on iterative sampling for both $\beta$ and $\sigma^2$ to estimate the assurance. We specify analysis priors $\beta\given\sigma^2 \sim N(\mu_{\beta}^{(a)}, \sigma^2V_{\beta}^{(a)})$ and $\sigma^{2} \sim IG(a^{(a)}, b^{(a)})$, where the superscripts $(a)$ signify analysis priors.

We had previously derived the posterior distribution of $\beta$ in Section~\ref{subsec: ohagan_stevens_linear_regression_known_sigma} expressed as $p(\beta\given y_n, \sigma^{2}) = N(\beta\given M_nm_n, \sigma^{2}M_n)$, where $M_n = (V_\beta^{-1(a)} + X^{\top} V_n^{-1} X)^{-1}$
and $m_n = V_\beta^{-1(a)}\mu_\beta^{(a)} + X^{\top}V_n^{-1}y_n$.
The posterior distribution of $\sigma^{2}$ is obtained by integrating out $\beta$ from the joint posterior distribution of $\{\beta, \sigma^2\}$, which yields
\begin{equation}\label{eq: postbeta}
   \begin{split}
   p(\sigma^2\given y_n) &\propto IG(\sigma^2\given a^{(a)},b^{(a)}) \times \int
   N(\beta \given \mu_{\beta}, \sigma^2 V_{\beta}) \times N(y_n \given X\beta,
   \sigma^2 V_{n}) d\beta\\ &\propto \left(\frac{1}{\sigma^2}\right)^{a^{(a)}+
   \frac{n}{2} + 1}\exp\left\{-\frac{1}{\sigma^2}\left(b^{(a)} +
   \frac{c^{\ast}}{2}\right)\right\}\;.
   \end{split}
\end{equation}
Therefore, $p(\sigma^2\given y_n) =
IG\left(\sigma^2\given a^{\ast}, b^{\ast}\right)$, where $a^{\ast} = a^{(a)} + \frac{n}{2}$ and $b^{\ast} = b^{(a)} + \frac{c^{\ast}}{2} = b^{(a)} + \frac{1}{2}
\left\{\mu_{\beta}^{\top (a)} V_{\beta}^{-1(a)}\mu_{\beta}^{(a)} +
y_n^{\top}V_n^{-1}y_n -  m_n^{\top}M_nm_n\right\}$.

Recall the design stage objective aims to identify sample size $n$ that
is needed to attain the assurance level specified by the investigator. Similar to
Section~\ref{subsec: ohagan_stevens_linear_regression_known_sigma} we
will need the marginal distribution of $y_n$ with priors placed on both $\beta$ and $\sigma^2$. We denote these design priors as $\beta^{(d)}$ and $\sigma^{2(d)}$, respectively, to signify that we are working within the design stage. Derivation steps are almost identical to those outlined in Equation~\eqref{eq: ohagan_stevens_linear_regression_design_priors} for the known $\sigma^2$ case. With $\sigma^{2(d)}$ now treated as an unknown parameter, the marginal distribution of $y_n$, given $\sigma^{2(d)}$, under the design prior is derived from $y_n =  X_n\beta^{(d)} + e_n$, $ e_n \sim N(0, \sigma^{2(d)} V_n)$, $\beta^{(d)} = \mu_{\beta}^{(d)} + \omega; \quad \omega \sim  N(0, \sigma^{2(d)} V_{\beta}^{(d)})$, where $\beta^{(d)}\sim N(\mu_{\beta}^{(d)},\sigma^{2(d)} V_{\beta}^{(d)})$
and $\sigma^{2(d)} \sim IG(a^{(d)}, b^{(d)})$. Substituting $\beta^{(d)}$ into
$y_n$ gives us $y_n = X_n\mu_{\beta}^{(d)} + (X_n\omega + e_n)$ such that $X_n\omega + e_n \sim N(0, \sigma^{2(d)}(V_n + X_n V_{\beta}^{(d)} X_n^{\top}))$.
The marginal distribution of $p(y_n\given \sigma^{2(d)})$ is
\begin{equation}\label{eq:marg_yn}
   y_n \given \sigma^{2(d)} \sim N(X_n \mu_{\beta}^{(d)}, \sigma^{2(d)} V_n^{*}); \quad
   V_n^{*} = X_n V_{\beta}^{(d)} X_n^{\top} + V_n.
\end{equation}
Equation~\eqref{eq:marg_yn} specifies our data generation model for ascertaining sample size.

The pseudocode in Algorithm~\ref{alg:unknown_var} evaluates Bayesian assurance. Each iteration comprises the design stage, where the data is generated, and an analysis stage where the data is analyzed to ascertain whether a decision favorable to the hypothesis has been made. In the design stage, we draw $\sigma^{2(d)}$ from
$IG(a_\sigma^{(d)}, b_\sigma^{(d)})$ and generate the data from our sampling distribution from \eqref{eq:marg_yn}, $y_n \sim N(X\mu_{\beta}^{(d)}, \sigma^{2(d)}(XV_{\beta}^{(d)}X^{\top} + V_n))$. For each such data set, $\{y_n, X_n\}$, we perform Bayesian inference for $\beta$ and $\sigma^2$. Here, we draw $J$ samples of $\beta$ and $\sigma^2$ from their respective posterior distributions and compute the proportion of  these $J$ samples that satisfy $u^{\top} \beta_j > C$. If the proportion exceeds a certain threshold $1-\alpha$, then the analysis objective is met for that dataset. The above steps for the design and analysis stage are repeated
for $R$ datasets and the proportion of the $R$ datasets that meet the analysis objective, i.e., deciding in favor of $H$, correspond to the Bayesian assurance.
                                                                                                                                                                                                                                                            \begin{spacing}{0.5}
\begin{algorithm}[htbp]
\caption{Bayesian assurance algorithm for unknown variance}
\label{alg:unknown_var}
\begin{algorithmic}[1]
\Procedure{bayes2}{$n$, $u$, $C$, $R$, $X$, $V_n$, $V_{\beta}^{(d)}$,
  $V_{\beta}^{-1(a)}$, $\mu_{\beta}^{(d)}$, $\mu_{\beta}^{(a)}$, $a^{(d)}$,
  $a^{(a)}$, $b^{(a)}$, $b^{(d)}$, $\alpha$}
\State{count1 = $0$} \Comment{counts iterations that meet analysis objective}\\
\For{$i$ in range $1:R$} \Comment{$R$ denotes number of generated datasets}
\State{\textbf{Begin Design Stage}}
\State{$\gamma^2 \gets $ IG$(a^{(d)}, b^{(d)})$}
\State{count2 = $0$}
\Comment{tracks meeting analysis objective for generated data}
\State{$y \gets$ $n \times 1$ vector sampled from MVN$(X\mu_{\beta}^{(d)}, \gamma^2 (XV_{\beta}^{(d)}X^{\top} + V_n))$}
\State{\textbf{End Design Stage}}\\
\State{\textbf{Begin Analysis Stage}}
\State{Compute the components that make up the posterior
  distributions of $\beta$ and $\sigma^2$:}
\State{$M \gets (V_\beta^{-1(a)} + X^{\top} V_n^{-1} X)^{-1}$}
\State{$m \gets V_\beta^{-1(a)}\mu_\beta^{(a)} + X^{\top}V_n^{-1}y$}
\State{$a^{\ast} = a^{(a)} + \frac{n}{2}$}
\State{$b^{\ast} = b^{(a)} + \frac{1}{2}\{\mu_\beta^{\top(a)}
  V_\beta^{-1(a)}\mu_\beta^{(a)} + y^{\top}V_n^{-1}y - m^{\top}Mm\}$}\\

\For{$j$ in range 1:J} \Comment{$J$ denotes number of  posterior samples}
\State{$\sigma^2 \gets$ IG($a^{\ast}, b^{\ast}$)}
\State{$\beta \gets$ $p \times 1$ vector sampled from MVN($Mm, \sigma^2 M$)}
\If{$u^{\top} \beta \leq C$}
\State{count2 $\gets$ count2 + 1}
\Else
\If{$u^{\top} \beta > C$}
\State{count2 $\gets$ count2}
\EndIf
\EndIf
\EndFor\\

\If{count2 / J $\leq \alpha$}
\State{count1 = count1 + 1}
\Else
\If{count2 / J $> \alpha$}
\State{count1 = count1}
\EndIf
\EndIf
\State{\textbf{End Analysis Stage}}\\
\EndFor
\State{\textbf{assurance} $\gets$ count1 / R}\\
\Return{assurance}\\
\EndProcedure
\end{algorithmic}
\end{algorithm}
\end{spacing}

\section{Two-Stage Paradigm Applications}\label{two_stage_apps_section}

The following sections explore three existing sample size determination approaches. We show how these approaches emerge as special cases of our framework with an appropriate formulation of analysis and design stage objectives. Assurance curves are produced via simulation and
pseudocodes of the algorithms are included.


\subsection{Sample Size Determination in Cost-Effectiveness Setting}\label{subsec:ohagan_stevens}

The first application selects a sample size based on
the cost-effectiveness of new treatments undergoing Phase 3 clinical trials \citep{ohagan}. As delineated in Section~\ref{BayesianSampleSizeSec}, we construct the two-stage paradigm under the context of a conjugate linear model and
generalize it to the case where the population variance $\sigma^2$ is unknown.
We cast the example in  \cite{ohagan} within our framework to assess overall performance and our ability to emulate the analysis in \cite{ohagan}.

Consider designing a randomized clinical trial where $n_1$ patients are administered
Treatment~1 and $n_2$ patients are administered Treatment~2 under some suitable model
and study objectives. Let $c_{ij}$ and $e_{ij}$ denote the observed cost and efficacy values, respectively, corresponding to patient $j=1,2,\ldots, n_i$ receiving treatment $i$ for $i = 1, 2$ treatments, where $n_i$ is the number
of patients in the $i$-th treatment group. Furthermore, the expected population mean cost efficacy under treatment $i$ are set  to be $E(c_{ij}) = \gamma_{j}$ and $E(c_{ij}) = \mu_{i}$, respectively. The
variances are taken to be
$Var(c_{ij}) = \tau_i^2$ and $Var(e_{ij}) = \sigma_i^2$.
For simplicity, we assume equal sample sizes within the treatment groups so that $n = n_1 = n_2$. We also assume equal sample variances for the costs and  efficacies such that $\tau^2$ = $\tau_1^2 = \tau_2^2 $ and $\sigma^2 = \sigma_1^2 = \sigma_2^2$.

\cite{ohagan} utilize the net monetary benefit measure,
\begin{equation}
\label{eq:netmon}
  \xi = K(\mu_2 - \mu_1) - (\gamma_2 - \gamma_1),
\end{equation}
where $\gamma_2 - \gamma_1$ and $\mu_2 - \mu_1$ denote the true differences in costs and efficacies, respectively, between Treatment 1 and Treatment 2, and $K$ represents the maximum price that a health care provider is willing to pay in order to obtain a unit increase in efficacy, also known as the threshold unit cost. The quantity $\xi$ acts as a measure of cost-effectiveness.

Since the net monetary benefit formula expressed in Equation~\eqref{eq:netmon}
involves assessing the cost and efficacy components conveyed within each
of the two treatment groups, we set $\beta = (\mu_1, \gamma_1, \mu_2, \gamma_2)^{\top}$,
where $\mu_i$ and $\gamma_i$ denote the efficacy and
cost for treatments $i = 1, 2$.
Next, we specify $y_n$ as a $4n \times 1$ vector consisting of $2\times 1$ vectors $y_{ij} = (c_{ij}, e_{ij})^{\top}$, $i=1,2$ and $j=1,2,\ldots,n$. Each individual observation
is allotted one row in the linear model. The design matrix $X_n$ is a $4n\times 4$ block diagonal with the $n\times 1$ vector of ones, $1_n$, as the blocks. With $n = n_1 = n_2$, $\sigma^2 = \sigma_1^2 = \sigma_2^2$ and $\tau^2 = \tau_1^2 = \tau_2^2$, our variance matrix is
$\displaystyle
\sigma^2\underset{4n \times 4n}{V_n} =
\sigma^2
\begin{pmatrix}
I_n & O & O & O\\
O & \frac{\tau^2}{\sigma^2}I_n & O & O\\
O & O & I_n & O\\
O & O & O & \frac{\tau^2}{\sigma^2}I_n\\
\end{pmatrix},
$
where $\sigma^2$ is factored out to comply with our conjugate
linear model formulation expressed in Equation~\eqref{eq: bayesian_conj_regression}.

In the analysis stage, we use the posterior distribution for $\beta$ if $\sigma^2$ is fixed or for $\{\beta, \sigma^2\}$ if $\sigma^2$ needs to be estimated; recall Sections~\ref{subsec: ohagan_stevens_linear_regression_known_sigma}~and~\ref{subsec:unknownvar}. The posterior distribution is needed only for the analysis stage, hence it is computed using the analysis priors. Since there is no data in the design stage, there is no posterior distribution. We use the design priors as specifications for the population from which the data is generated. That is, the design priors yield the sampling distribution $y_n\given \sigma^{2(d)}$.
\cite{ohagan} define $\mu_{\beta}^{(d)} = (5, 6000, 6.5, 7200)^{\top}$ and $\displaystyle V_{\beta}^{(d)} =
    \begin{pmatrix}
  4 & 0 & 3 & 0\\
  0 & 10^7 & 0 & 0\\
  3 & 0 & 4 & 0\\
  0 & 0 & 0 & 10^7.
  \end{pmatrix}$.
We factor out $\sigma^2$ from $V_{\beta}^{(d)}$ to be consistent with the conjugate Bayesian formulation in Section~\ref{subsec: ohagan_stevens_linear_regression_known_sigma} so $\displaystyle \sigma^2 V_{\beta}^{(d)} =
      \sigma^2
    \begin{pmatrix}
    4/\sigma^2 & 0 & 3 & 0\\
    0 & 10^7/\sigma^2 & 0 & 0\\
    3 & 0 & 4/\sigma^2 & 0 \\
    0 & 0 & 0 & 10^7/\sigma^2
    \end{pmatrix}$. Lastly, we set the posterior probability of deciding in favor of $H$ to at least $0.975$, which is equivalent to a Type-I error of $\alpha = 0.025$ in frequentist two-sided hypothesis tests.

\subsection{Design for Cost-Effectiveness Analysis}

Consider designing a trial to evaluate the cost effectiveness of a new treatment with an original treatment. We seek the tenability of $H: \xi > 0$, where $\xi$ is the net monetary benefit defined in Equation~\eqref{eq:netmon}. Using the inputs specified in Section~\ref{subsec:ohagan_stevens} we execute simulations in the known and unknown $\sigma^2$ cases to emulate \cite{ohagan} as a special case.

\subsubsection{Simulation Results in the Known $\sigma^2$ Case}\label{sim_known}

\pari Table~\ref{tab:sim_results} presents Bayesian assurance values corresponding to different values of $K$ and sample size $n$. The ``maxiter'' variable, as described in Algorithm~\ref{alg:known_var}, is the number of data sets being simulated. All of the resulting assurance values in Table~\ref{tab:sim_results} for all combinations of $K$ and $n$ are close to 0.70. Looking by columns, we see that the assurance values exhibit minor deviations in both directions for all cases as we increase the number of iterations being implemented in each run. No obvious trends of precision are showcased in any of the
four cases. Looking across rows we observe that larger sample sizes tend to yield assurance values that are consistently closer to the 0.70 mark, which is to be expected. The first column, corresponding to the case with the largest sample size of $n = 1048$, consistently produced results that meet the assurance criteria of 0.70. These results show that sampling from the posterior provides results very similar to those reported in \cite{ohagan}.

  \begin{table}[t]
  \begin{center}
  \begin{tabular}{c c c c c c}
  \multicolumn{6}{c}{Outputs from Bayesian Assurance Algorithm} \\
  \hline
  maxiter & \footnotesize{\shortstack{K =5000 \\ n = 1048}}
  & \footnotesize{\shortstack{K = 7000 \\ n = 541}} & \footnotesize{\shortstack{K =   10000 \\ n = 382}}
  & \footnotesize{\shortstack{K = 20000 \\ n = 285}}\\
  \hline
  250 & 0.708 & 0.676 & 0.688 & 0.716 \\
  \hline
  500 & 0.701 & 0.714 & 0.676 & 0.698\\
  \hline
  1000 & 0.700 & 0.694 & 0.697 & 0.719\\
  \end{tabular}
  \end{center}
  \caption{Simulation results from our Bayesian assurance algorithm with
    varying number of iterations in each of the four cost-effectiveness cases.}
  \label{tab:sim_results}
  \end{table}

\begin{figure}[t]
  \centering
  \includegraphics[width=0.5\textwidth]{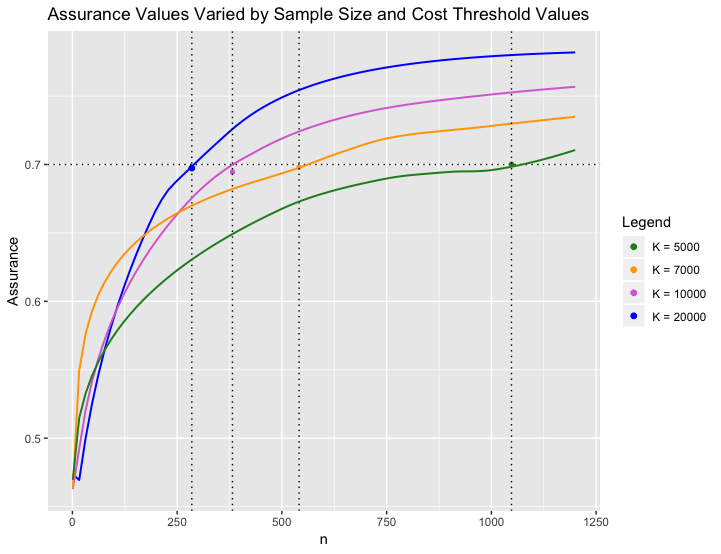}
  \caption{Assurance curves for each of the four different unit
    cost threshold cases $K$. Dotted vertical lines indicating
    when a presumed 0.70 assurance level is achieved according to
    O'Hagan and Stevens (from left to right):
    n = 285, n = 382, n = 541, and n = 1048.}
  \label{fig:combined_fixedcase}
\end{figure}

As a supplement to
Table~\ref{tab:sim_results},
it is also helpful to feature a visual representation of the relationship between sample size and assurance. The second part of our assessment compares the assurance curves obtained from multiple runs of our Bayesian simulation function
with varying sample sizes $n$. This was done for each of the four unit threshold cost assignments. A combined
plot incorporating all four assurance curves is shown in
Figure~\ref{fig:combined_fixedcase} and a sample of the exact
assurance values computed can be found in Table~\ref{tab:sim_numbers}. A smoothing feature from the $\textbf{ggplot2}$ package in R is implemented, which
fits the observed assurance points to a $\log(x)$ function.  We explicitly mark the four assurance points that our
algorithm returned with sample sizes $n$ and cost threshold values $K$ that correspond to reported assurance levels of 0.70 (horizontal line) in \cite{ohagan}.

\begin{figure}[t]
  \centering
    \includegraphics[width=0.45\textwidth]{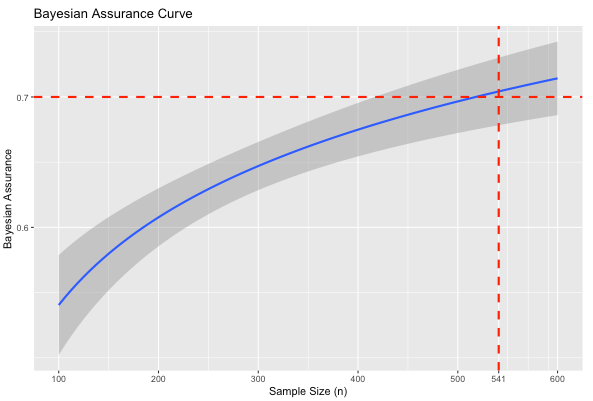}
    \includegraphics[width=0.45\textwidth]{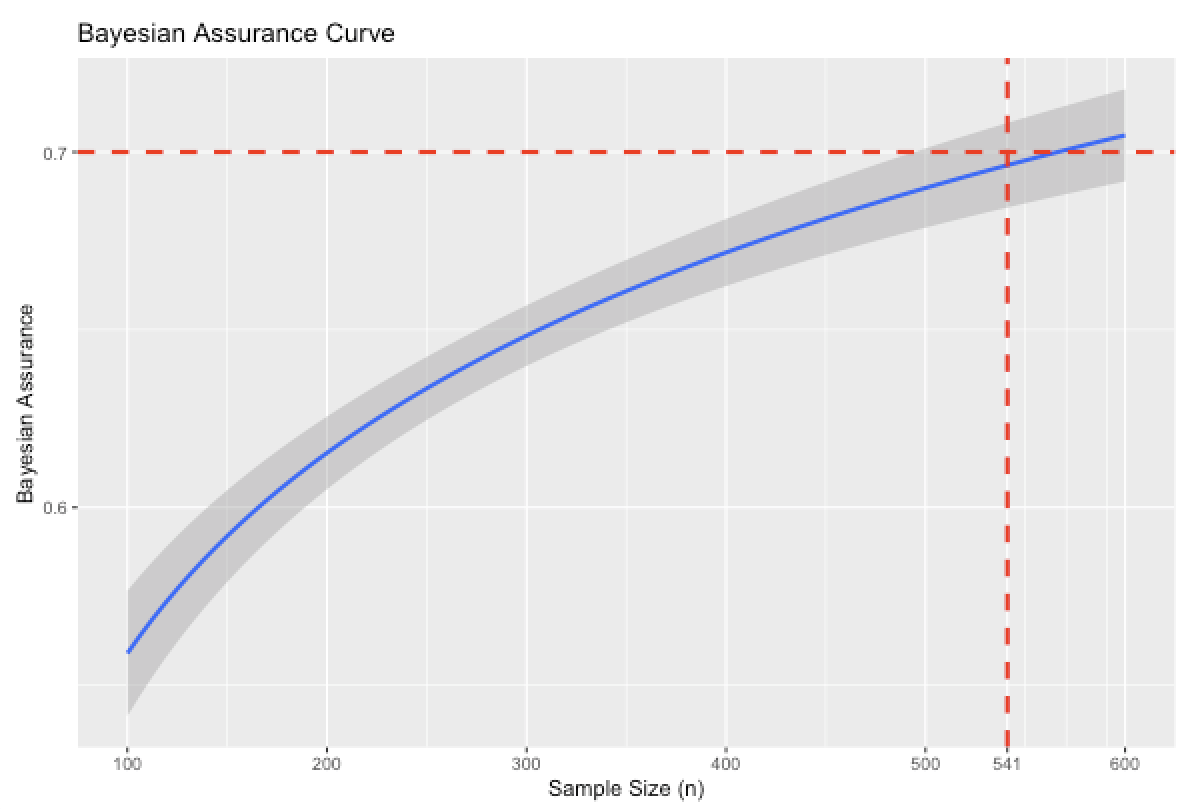}
    \caption{Resulting Bayesian assurance curves obtained from two separate
    simulation runs for the case with weak analysis priors and
    cost threshold value K = 7000.}
  \label{fig:sidebyside}
\end{figure}

\begin{table}[t]
\centering
\caption{Selected subset of assurance values computed by algorithm for fixed variance case (note that many points were
omitted here simply to make table more concise).
Pairs highlighted in red correspond to the 4 points that were explicitly plotted in Figure~\ref{fig:combined_fixedcase}.}
\label{tab:sim_numbers}
    \begin{tabular}{ C{11mm}C{16mm} C{11mm}C{16mm} C{11mm}C{16mm} C{11mm}C{16mm}}
    \toprule
         \multicolumn{2}{c}{K = 20000}
                            &   \multicolumn{2}{c}{K = 10000} & \multicolumn{2}{c}{K = 7000} & \multicolumn{2}{c}{K = 5000}\\
    \cmidrule(lr){1-2}\cmidrule(lr){3-4}\cmidrule(lr){5-6}\cmidrule(lr){7-8}
        n & Assurance & n & Assurance & n & Assurance & n & Assurance\\
    \midrule
   1 & 0.473 & 1 & 0.463 & 1 & 0.463 & 1 & 0.470 \\
  185 & 0.655 & 282 & 0.673 & 440 & 0.687 & 500 & 0.667 \\
  205 & 0.669 & 332 & 0.693 & 490 & 0.694 & 750 & 0.689 \\
  235 & 0.687 & \textbf{\leavevmode\color{red}382} & \textbf{\leavevmode\color{red}0.695}& \textbf{\leavevmode\color{red}541} & \textbf{\leavevmode\color{red}0.698} & 875 & 0.699 \\
  \textbf{\leavevmode\color{red}285} & \textbf{\leavevmode\color{red}0.697} & 482 & 0.716 & 640 & 0.712 & 1000 & 0.695 \\
  335 & 0.716 & 750 & 0.743 & 750 & 0.719 & \textbf{\leavevmode\color{red}1048} &\textbf{\leavevmode\color{red}0.700}\\
  1200 & 0.782 & 1200 & 0.758 & 1200 & 0.735 & 1200 & 0.710 \\
    \bottomrule
    \end{tabular}
\end{table}

We also provide a separate graphical display showcasing how the assurance behaves in separate runs to assess consistency in results.
Figure~\ref{fig:sidebyside} provides a side by side comparison of the assurance curves for the case, where the unit threshold cost is $K=7000$. The dotted red lines correspond to the 0.7
threshold and the effects of simulation errors can be seen
through the slight differences in results between the two implementations. More specifically, the left image indicates that we achieve our 0.70 assurance level slightly before \cite{ohagan} reported a
sample size of $n = 541$, whereas the right image shows that the assurance is
achieved slightly after $n = 541$. Such minor fluctuations are to be expected due to Monte Carlo errors in simulation.

\subsubsection{Simulation Results in Unknown $\sigma^2$ Case}\label{sim_unknown}
\pari We now consider the setting where $\sigma^2$ is unknown. This extends the analysis in \cite{ohagan} who treated the cost-effectiveness problem with fixed variances. Table~\ref{tab:sim_results2} does not align as closely as the assurance results we had obtained from implementing the fixed $\sigma^2$ simulation in Section~\ref{sim_known}.

Referring to Table~\ref{tab:sim_results2}, we let $R$ be the number
of outer loop iterations. The primary purpose of the outer loop is to randomly draw design stage variances  $\sigma^{2(d)}$ from the $IG(a^{(d)}, b^{(d)})$ distribution.
Recall from Section~\ref{subsec:unknownvar}
that $\sigma^{2(d)}$ is used for computing the variance of the marginal distribution
from which we are drawing our sampled observations, $y\given \sigma^{2(d)}$.
The inner-loop iterations sample data using the marginal distribution of $\sigma^2$ from Equation~\eqref{eq: postbeta}. The number of iterations in the inner-loop is set to 750 for all cases. We notice that a majority of our trials report assurance values close to the 0.70 mark, particularly for the case in which we set the sample size to $n = 1048$. The trial that exhibited the greatest deviation was for threshold cost $K = 20000$ with corresponding sample size  $n = 285$, which returned an assurance of 0.58. This is most likely attributed to using a smaller sample size to gauge the effect size.

\begin{table}[t]
    \begin{center}
    \begin{tabular}{ c c c c c c}
    \multicolumn{6}{c}{Outputs from Bayesian Assurance Algorithm} \\
    \hline
    R & \footnotesize{\shortstack{K =5000 \\ n = 1048}}
    & \footnotesize{\shortstack{K = 7000 \\ n = 541}} & \footnotesize{\shortstack{K = 10000 \\ n = 382}}
    & \footnotesize{\shortstack{K = 20000 \\ n = 285}}\\
    \hline
    100 &  0.698 & 0.718 & 0.72 & 0.601 \\
    \hline
    150 & 0.702 & 0.713 & 0.72 & 0.579
    \end{tabular}
    \end{center}
    \caption{Simulation results from the nonfixed Bayesian assurance algorithm with
      varying number of iterations in each of the four cost-effectiveness cases.}
    \label{tab:sim_results2}
    \end{table}

\begin{figure}[h]
    \centering
    \includegraphics[width=0.6\textwidth]{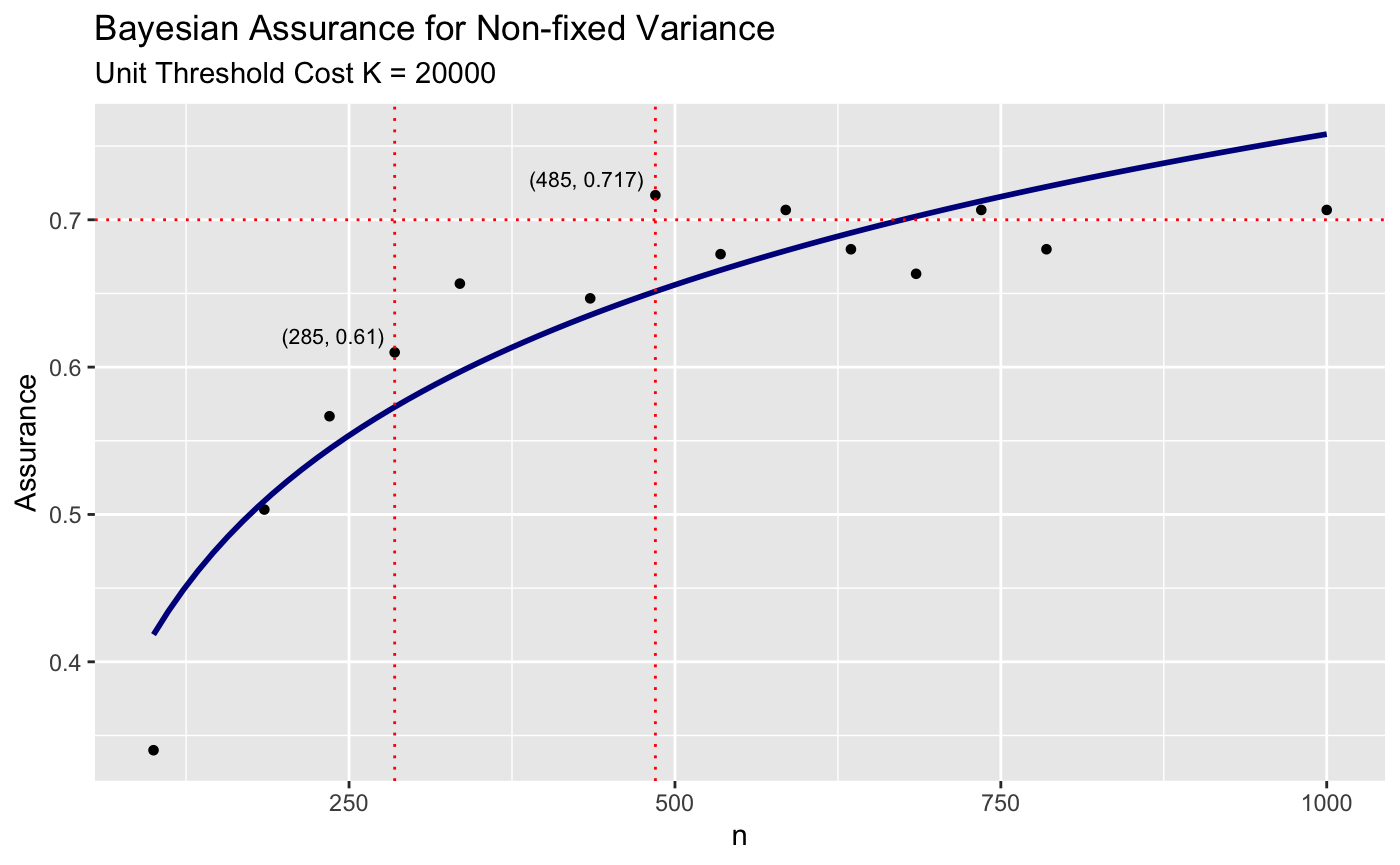}
    \caption{Assurance curve based on results of algorithm corresponding
      to unknown variance.}
    \label{fig:nonfixedvar}
    \end{figure}

    A visual depiction for this case can be seen in Figure~\ref{fig:nonfixedvar}.
    The dashed line on the left showcases the expected minimum sample size
    needed to achieve a 0.70 assurance whereas the dashed line on the right
    marks the point at which our algorithm actually achieves this desired threshold.
    The reality of the situation is that the problem setup gets changed quite a bit
    once we remove the assumption that $\sigma^2$ is known and fixed.
    If we look at the individual points marked on the plot,
    assurance values of 0.61 and 0.71
    don't appear too different. If we were to solely account for the fact that these
Monte Carlo estimates are subject to error given that the estimates
are based on means and variances that were
compositely sampled rather than being taken in as fixed assignments,
our algorithm performs remarkably well, but there are still points
to be wary about.

The x-axis of the plot indicates that an assurance of 0.70 (red dotted line) can only
be ensured once we recruit a sample size of at least $n = 485$ per treatment group. This is substantially larger compared to the known $\sigma^2$ case, suggesting
a need to recruit nearly twice as many participants as what was needed in Table~\ref{tab:sim_results}. These results evince the pronounced impact of uncertainty in the design on the sample size needed to achieve a fixed level of Bayesian assurance.

\subsection{Sample Size Determination with Precision-Based Conditions}\label{subsec:precision}
\pari We now consider a few alternate Bayesian approaches for sample size determination  and demonstrate how these methods can be embedded within the two-stage Bayesian framework. We also identify special cases that overlap with the frequentist setting.

\cite{cj} constructs rules based on a fixed precision level $d$. In the frequentist setting, if $X_i \sim N(\theta, \sigma^2)$ for $i = 1,..., n$ observations and variance $\sigma^2$ is known, the precision can be
calculated using $d = z_{1-\alpha/2}\frac{\sigma}{\sqrt{n}}$, where $z_{1-\alpha/2}$ is the critical value for the $100(1-\alpha/2)\%$ quartile of the standard normal distribution. Simple rearranging leads to following expression for sample size,
\begin{equation}
  \label{eq:samp_size_d}
  n = z_{1-\alpha/2}^2\frac{\sigma^2}{d^2}\; .
\end{equation}
Given a random sample with mean $\bar{x}$, suppose the goal is to estimate  population mean $\theta$. The analysis objective entails deciding whether or not the absolute difference between $\bar{x}$ and $\theta$ falls within a margin of error no greater than $d$. Given data $\bar{x}$ and a pre-specified confidence level $\alpha$,
the assurance can be formally expressed as
\begin{equation} \label{eq: prec_cond}
\delta = P_{\bar{x}}\{\bar{x}: P(|\bar{x} - \theta| \leq d) \geq 1-\alpha\}\;.
\end{equation}

To formulate the problem in the Bayesian setting, suppose $x_1, \cdots, x_n$ is a random sample from $N(\theta, \sigma^2)$
and the sample mean is distributed as $\bar{x}|\theta, \sigma^2 \sim N(\theta, \sigma^2/n)$.

We assign $\theta \sim N(\theta_0^{(a)}, \sigma^2/n_a)$ as the analysis prior, where $n_a$ quantifies the amount of prior information we have for $\theta$. Adhering to the notation in previous sections, subscript $(a)$ denotes we are working within the analysis stage. Referring to Equation~\eqref{eq: prec_cond}, the analysis stage objective
is to observe if the condition, $|\bar{x} - \theta| \leq d$,
is met. Recall that if the analysis objective holds to
a specified probability level, then the corresponding sample size of the data being passed through the condition  is sufficient in fulfilling the desired precision level for the study.

Additional steps can be taken to further expand out Equation~\eqref{eq: prec_cond}.
The posterior of $\theta$ can be obtained by taking the product of the prior and likelihood, giving us
\begin{equation}
  \label{eq:post_theta_adcock}
  N\Bigg(\bar{x} {\left| \theta, \frac{\sigma^2}{n}\right.}\Bigg) \times
  N\Bigg(\theta {\left | \theta_0^{(a)}, \frac{\sigma^2}{n_a}\right.}\Bigg)
  = N\Bigg(\theta {\left | \lambda, \frac{\sigma^2}{n_a+n}\right.}\Bigg),
 \end{equation}
where $\lambda = \frac{n\bar{x}+n_a\theta_0^{(a)}}{n_a+n}$.
From here we can further evaluate the condition using parameters from the posterior of $\theta$ to obtain a more explicit version of the analysis stage objective. Starting from
$P(|\bar{x} - \theta| \leq d) = P(\bar{x} - d \leq \theta \leq \bar{x} + d)$, we can
standardize all components of the inequality using the posterior parameter
values of $\theta$, leading us to
\begin{multline*}
P(|\bar{x} - \theta| \leq d) = P\left(\frac{\bar{x} - d - \lambda}{\sigma/\sqrt{n_a + n}}
\leq \frac{\theta - \lambda}{\sigma/\sqrt{n_a + n}} \leq \frac{\bar{x} + d - \lambda}{\sigma/\sqrt{n_a + n}}\right)
\\ = P\left( \frac{\bar{x} - d - \lambda}{\sigma/\sqrt{n_a + n}}
\leq Z \leq \frac{\bar{x} + d - \lambda}{\sigma/\sqrt{n_a + n}} \right).
\end{multline*}
Simplifying the result gives us
\begin{equation}
  \label{eq:adcock_region_simplified}
  \delta = \left\{ \bar{x}: \Phi\left[\frac{\sqrt{n_a + n}}{\sigma}
  (\bar{x} + d - \lambda)\right] - \Phi\left[\frac{\sqrt{n_a+n}}{\sigma}(\bar{x}
  - d - \lambda)\right] \geq 1-\alpha \right\}.
\end{equation}

Moving on to the design stage, we need to construct a protocol for sampling data that will be used to evaluate the analysis objective. This is achieved
by first setting a separate design stage prior on $\theta$ such that $\theta \sim N(\theta_0^{(d)}, \sigma^2/n_d)$, where $n_d$ quantifies
our degree of belief towards the population from which the sample will be drawn. Given that $\bar{x}|\theta, \sigma^2 \sim N(\theta, \sigma^2/n)$, the marginal distribution of $\bar{x}$ can be computed using straightforward substitution based on $\bar{x} = \theta + \epsilon;$ $\epsilon \sim N(0, \sigma^2/n)$ and
$\theta = \theta_0^{(d)} + \omega;$ $\quad \omega \sim N(0, \sigma^2/n_d)$.  Substitution $\theta$ into the expression for $\bar{x}$ gives us
$\bar{x}=\theta_0^{(d)} + (\omega + \epsilon);$
$(\omega + \epsilon) \sim N\Big(0, \frac{\sigma^2(n_d+n)}{n_dn}\Big)
= N(0, \sigma^2/p)$, where $1/p = 1/n_d + 1/n$. The marginal
of $\bar{x}$ is therefore $N(\bar{x}|\theta_0^{(d)}, \sigma^2/p)$, where we will be
iteratively drawing our samples from to check if the sample means satisfy the condition
derived in Equation~\eqref{eq:adcock_region_simplified}. Algorithm~\ref{alg:adcock}
provides a skeleton of the code that was used to implement our simulations.


\begin{algorithm}
    \caption{Bayesian assurance algorithm using Adcock's condition for known variance in the univariate case}
  \label{alg:adcock}
  \begin{algorithmic}[1] 
  \Procedure{bayes adcock}{$n$, $d$, $\theta_0^{(a)}$, $\theta_0^{(d)}$, $n_a$, $n_d$,
    $\sigma^2$, $\alpha$}

  \State{count = $0$} \Comment{keeps track of the iterations that satisfy the analysis obj}\\
  \State{maxiter = $1000$} \Comment{arbitrary number of iterations to loop thru}

  \For{$i$ in range $1$ : maxiter}
  \State{\textbf{Design Stage Starts}}
  \State{$\sigma^2_d \gets \sigma^2 \frac{n_d + n}{n n_d}$}
  \State{$\bar{x} \gets$ single value generated from $N(\theta_0^{(d)}, \sigma^2_d)$}
  \State{\textbf{Design Stage Ends}}\\

  \State{\textbf{Analysis Stage Starts}}
  \Comment{Computes components of the posterior distribution of parameter $\theta$:}

  \State{$\lambda \gets \frac{n_a \theta_0^{(a)} + n\bar{x}}{n_a + n}$}
  \State{$\sigma^2_a \gets \frac{\sigma^2}{n_a + n}$}
  \State{$\theta \gets$ single value generated from $N(\lambda, \sigma^2_a)$}\\

  \State{$\phi_1 \gets \frac{\sqrt{n_a + n}}{\sigma}(\theta + d - \lambda)$}
  \State{$\phi_2 \gets \frac{\sqrt{n_a + n}}{\sigma}(\theta - d - \lambda)$}\\

  \If{$\Phi(\phi_1) - \Phi(\phi_2) \geq 1 - \alpha$}
  \State{$Z_i \gets 1$}

  \Else
  \If{$\Phi(\phi_1) - \Phi(\phi_2) < 1 - \alpha$}
  \State{$Z_i \gets 0$}
  \EndIf
  \EndIf\\

  \State{count $\gets$ count + $Z_i$}
  \State{\textbf{Analysis Stage Ends}}
  \EndFor\\

  \State{\textbf{assurance} $\gets$ count / maxiter}\\
  \Return{assurance}\\
  \EndProcedure
  \end{algorithmic}
  \end{algorithm}

\subsubsection{Convergence to the Frequentist Setting}
\pari Unlike the cost-effectiveness application, the
precision-based setting in \cite{cj} is not situated in a hypothesis testing framework. Hence, we cannot compute a set of corresponding power values that are directly comparable to our simulated assurance values.
Nevertheless, an appropriate formulation of the analysis and design stage precision parameters, $n_a$ and
$n_d$, can emulate this setting.

Referring to the derived expression for assurance in
Equation~\eqref{eq:adcock_region_simplified}, note that we are ultimately assessing whether the expression on the left hand side exceeds $1-\alpha$.
Using the frequentist sample size formula given in
Equation~\eqref{eq:samp_size_d}, we can work backwards from the sample size formula such that $1-\alpha$ is isolated on the right hand side of the inequality. We can then
compare the expressions to compare the behaviors in relation  to the probability of meeting the pre-specified condition.
Starting from Equation~\eqref{eq:samp_size_d}, simple rearrangement reveals
\[
n \geq z^2_{1-\alpha/2}\frac{\sigma^2}{d^2} \implies \frac{\sqrt{n}}{\sigma}d \geq z_{1-\alpha/2}
\implies \Phi\Bigg[\frac{\sqrt{n}}{\sigma}d\Bigg] \geq 1 - \alpha/2 \implies
2\Phi\left[\frac{\sqrt{n}}{\sigma}d\right] - 1 \geq 1 - \alpha.
\]
If we refer back to Equation~\eqref{eq:adcock_region_simplified},
it becomes clear that setting $n_a = 0$ will simplify the expression down to the same
expression we had previously obtained for the above frequentist scenario. Hence,
\begin{align*}
&\delta = \left\{ \bar{x}: \Phi\left[\frac{\sqrt{n_a + n}}{\sigma}
  (\bar{x} + d - \lambda)\right] - \Phi\left[\frac{\sqrt{n_a+n}}{\sigma}(\bar{x}
  - d - \lambda)\right] \geq 1-\alpha \right\} \\
  &\xRightarrow[]{n_a = 0} \left\{ \bar{x}: \Phi\left[\frac{\sqrt{n}}{\sigma}d\right] -
  \Phi \left[-\frac{\sqrt{n}}{\sigma}d\right] \geq 1 - \alpha \right\} = \left\{
  \bar{x}: 2\Phi \left[\frac{\sqrt{n}}{\sigma}d\right] - 1 \geq 1 - \alpha \right\}.
\end{align*}
In other words, if we let $\theta$ take on a weak analysis prior, we revert back to the frequentist setting in the analysis stage.

\subsubsection{Simulation Results under Precision-Based Conditions}
\begin{figure}[h]
  \centering
    \includegraphics[width=0.75\textwidth]{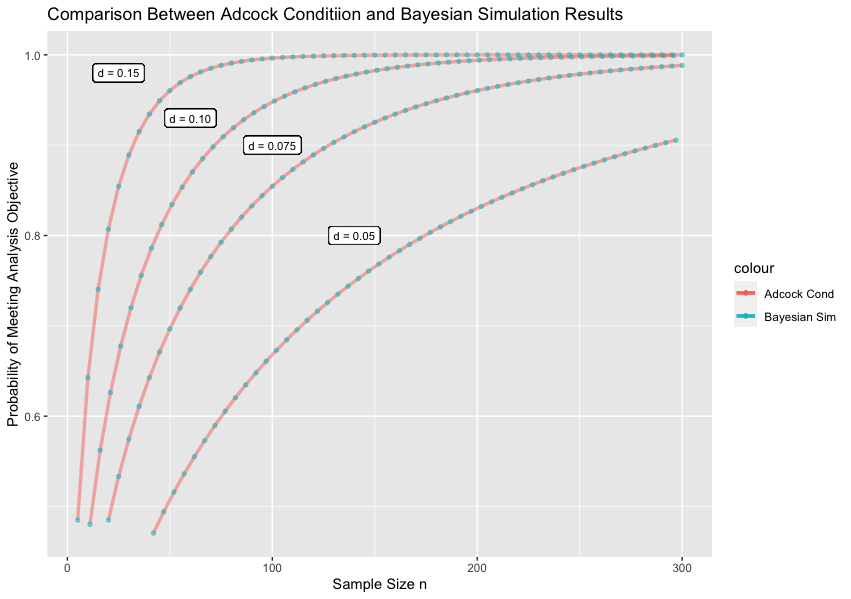}
    \caption{Overlay of simulated results and frequentist
    results given a weak analysis prior such that $n_a \rightarrow 0$.}
  \label{fig:adcocksim}
\end{figure}

We test our algorithm using different fixed precision parameters $d$
with varying sample sizes $n$. The remaining fixed parameters including $\sigma^2$,
$\theta_0^{(a)}$, and $\theta_0^{(d)}$ are randomly drawn from
the uniform distribution Unif(0, 1) for simplicity sake.
Figure~\ref{fig:adcocksim} displays the results of the Bayesian-simulated
points (marked in blue) in the case where weak analysis stage
priors were assigned overlayed on top of the frequentist results (marked
in red). Note that the Bayesian-simulated points denote the probability
of observing that the posterior of $\theta$ differing from the sample mean $\bar{x}$
within a range of $\bar{x} \pm d$ exceeds $1 - \alpha$.
From a general standpoint, these probabilities are obtained by
iterating through multiple samples
of size $n$ and observing the proportion of those samples that meet
the analysis stage objective from Equation~\eqref{eq:adcock_region_simplified}.
As we have shown in the previous section, this becomes trivial in the case where weak analysis priors are assigned as
we are left with a condition that is independent of $\bar{x}$. Hence,
we are able to obtain the exact same probability values as those
obtained from the frequentist formula. As shown in the plot, this can be
seen across all sample sizes $n$ for all precision parameters $d$.

\subsection{Sample Size Determination in a Beta-Binomial Setting}
We 
revisit the
hypothesis testing framework 
with %
proportions.
\cite{pham} outlines steps for determining exact sample sizes needed in estimating differences of two proportions in a Bayesian context. Let $p_i, i = 1, 2$ denote two independent proportions. In the frequentist setting, suppose the hypothesis test
to undergo evaluation is
$H_0: p_1 - p_2 = 0$ vs. $H_a: p_1 - p_2 \neq 0$. As described in \cite{pham},
one method of approach is to check whether or not $0$ is contained within
the confidence interval bounds of the true difference in proportions given by $\displaystyle (\hat{p_1} - \hat{p_2}) \pm z_{1-\alpha/2}(SE(\hat{p_1})^2 + SE(\hat{p_2})^2)^{1/2}$,
where $z_{1-\alpha/2}$ denotes the critical region, and $SE(\hat{p_i})$
denotes the standard error of $p_i$ obtained by
$SE(\hat{p_i}) = \sqrt{\frac{\hat{p_i}(1-\hat{p_i})}{n_i}}$. An interval without
$0$ contained within the bounds suggests there exists a significant difference between
the two proportions.

The Beta distribution is often used to represent outcomes tied to a
family of probabilities.
The Bayesian setting uses posterior credible intervals as an analog to the
frequentist confidence interval approach. As outlined in \cite{pham},
two individual priors are assigned to $p_1$ and $p_2$ such that
$p_i \sim Beta(\alpha_i, \beta_i)$ for $i = 1,2$.
In the case of binomial sampling, $X$ is treated as a random variable
taking on values $x = 0, 1,..., n$ to denote the number of favorable
outcomes out of $n$ trials. The proportion of favorable outcomes is
therefore $p = x/n$. Suppose a Beta prior is assigned to $p$
such that  $p \sim Beta(\alpha, \beta)$. The prior mean and variance are respectively
$\mu_{prior} = \frac{\alpha}{\alpha + \beta}$ and $\sigma^2_{prior} =
\frac{\alpha\beta}{(\alpha + \beta)^2(\alpha + \beta + 1)}$. Conveniently, given
that $p$ is assigned a Beta prior, the posterior of $p$ also takes on a Beta
distribution with mean and variance
\begin{align}
  \label{eq: post_binomial}
  \begin{split}
    \mu_{posterior} &= \frac{\alpha + x}{\alpha + \beta + n}\\
    \sigma^2_{posterior} &= \frac{(\alpha + x)(\beta + n - x)}{(\alpha + \beta + n)^2(\alpha + \beta + n + 1)}.
  \end{split}
\end{align}

Within the analysis stage, we assign two beta priors for $p_1$ and $p_2$ such that
$p_i \sim Beta(\alpha_i, \beta_i), i = 1, 2$. If we let $p_{d} = p_1 - p_2$ and $p_{\text{post}}$ and $\text{var}(p)_{\text{post}}$
respectively denote the posterior mean and variance of $p_{d}$,
it is straightforward to deduce that $p_{\text{post}} = \frac{\alpha_1 + x_1}{\alpha_1 + \beta_1 + n_1} -
\frac{\alpha_2 + x_2}{\alpha_2 + \beta_2 + n_2}$ and $\text{var}(p)_{\text{post}} =
\frac{(\alpha_1 + x_1)(\beta_1 + n_1 - x_1)}{(\alpha_1 + \beta_1 + n_1)^2(\alpha_1 + \beta_1 + n_1 + 1)} +
\frac{(\alpha_2 + x_2)(\beta_2 + n_2 - x_2)}{(\alpha_2 + \beta_2 + n_2)^2(\alpha_2 + \beta_2 + n_2 + 1)}$
from Equation~\eqref{eq: post_binomial}. Hence the resulting $100(1-\alpha)\%$
credible interval equates to $p_{\text{post}} \pm z_{1-\alpha/2}
\sqrt{\text{var}(p)_{\text{post}}}$, which, similar
to the frequentist setting, would be used to check whether $0$ is contained within the
credible interval bands as part of our inference procedure. This translates to
become our analysis objective, where we are interested in assessing
if each iterated sample outputs a credible interval that does not contain $0$.
We can denote this region of interest as $R(p)$ such that
\begin{equation}
\label{eq: region_cond}
R(p) = \left\{p: 0 \not\in
\left(p_{\text{post}} - z_{1-\alpha/2} \sqrt{\text{var}(p)_{\text{post}}}, \quad p_{\text{post}} +
z_{1-\alpha/2} \sqrt{\text{var}(p)_{\text{post}}}\right)\right\}.
\end{equation}
It follows that the corresponding assurance for assessing a significant
difference in proportions can be computed as
\[
\delta = P \left\{p_d: 0 \not\in
\left(p_{\text{post}} - z_{1-\alpha/2} \sqrt{\text{var}(p)_{\text{post}}},
\quad p_{\text{post}} +
z_{1-\alpha/2} \sqrt{\text{var}(p)_{\text{post}}}\right) \geq 1-\alpha \right\}.
\]

Moving on to the design stage, note that the simulated data in Beta-Binomial
setting pertains to the frequency of positive outcomes, $x_1$ and $x_2$,
observed among the two samples.
These frequency counts are observed from samples of size $n_1$ and $n_2$
based on given probabilities, $p_1$ and $p_2$, that are passed in the
analysis stage. Once
$p_1$ and $p_2$ are assigned, $x_1$ and $x_2$ values are randomly generated from
their corresponding binomial distributions, where $x_i \sim Bin(n_i, p_i), i = 1,2$. The
posterior credible intervals are subsequently computed to undergo assessment in
the analysis stage. These steps are repeated iteratively starting from the generation of
$x_1$ and $x_2$ values. The proportion of iterations with results that fall within
the region of interest expressed in Equation~\eqref{eq: region_cond}
equates to the assurance.
Algorithm~\ref{alg:phamgia}
provides a skeleton of the code used to implement our simulations.

\begin{algorithm}
    \caption{Bayesian assurance algorithm for difference in two independent proportions under Pham-Gia's credible interval condition}
\label{alg:phamgia}
\begin{algorithmic}[1] 
\Procedure{bayes pham-gia}{$n_1$, $n_2$, $p_1 = \text{NULL}$, $p_2 = \text{NULL}$, $\alpha_1$, $\alpha_2$, $\beta_1$, $\beta_2$,
$\alpha$}

\If{$\psi = 1$}
\State{$p_1 \gets$ single value generated from $\text{Unif}[p_1, p_1]$}
\State{$p_2 \gets$ single value generated from $\text{Unif}[p_2, p_2]$}
\ElsIf {$\psi = 0$}
\State{$p_1 \gets$ $\text{Beta}(\alpha_1, \beta_1)$}
\State{$p_2 \gets$ $\text{Beta}(\alpha_2, \beta_2)$}
\EndIf \\

\State{count = $0$} \Comment{keeps track of the iterations that satisfy the analysis obj}
\State{maxiter = $1000$} \Comment{arbitrary number of iterations to loop thru}\\

\For{$i$ in range $1$ : maxiter}
\State{\textbf{Design Stage Starts}}
\State{$x_1 \gets$ single value generated from $Bin(n_1, p_1)$}
\State{$x_2 \gets$ single value generated from $Bin(n_2, p_2)$}
\State{\textbf{Design Stage Ends}}\\

\State{\textbf{Analysis Stage Starts}}

\State{$p_{\text{post}} = \frac{\alpha_1 + x_1}{\alpha_1 + \beta_1 + n_1} - \frac{\alpha_2 + x_2}{\alpha_2 + \beta_2 + n_2}$}
\Comment{Computes posterior parameters of $p = p_1 - p_2$}
\State{$\text{var}(p)_{\text{post}} = \frac{(\alpha_1 + x_1)(\beta_1 + n_1 - x_1)}
{(\alpha_1 + \beta_1 + n_1)^2(\alpha_1 + \beta_1 + n_1 + 1)} +
  \frac{(\alpha_2 + x_2)(\beta_2 + n_2 - x_2)}{(\alpha_2 + \beta_2 + n_2)^2(\alpha_2 + \beta_2 + n_2 + 1)}$}\\

\State{lb = $p_{\text{post}} - z_{1-\alpha/2}\sqrt{var(p)_{\text{post}}}$} \Comment{Computes upper and lower bounds}
\State{ub = $p_{\text{post}} + z_{1-\alpha/2}\sqrt{var(p)_{\text{post}}}$}\\

\If{$0 < $ lb or $0 > $ ub}
\State{count $\gets$ count + $1$}
\EndIf
\State{\textbf{Analysis Stage Ends}}

\EndFor\\

\State{\textbf{assurance} $\gets$ count / maxiter}\\
\Return{assurance}\\
\EndProcedure
\end{algorithmic}
\end{algorithm}

\subsubsection{Relation to Frequentist Setting}
It is worth pointing out that there are no precision parameters to quantify the amount of information
we have on the priors being assigned. Directly showcasing parallel
behaviors between Bayesian and frequentist settings involve knowing the probabilities beforehand and passing them in as
arguments into the simulation.
Specifically, if $p_1$ and $p_2$ are known beforehand, we can express these
``exact'' priors as Uniform distributions such that $p_i \sim \text{U}[p_i, p_i]$.
We can then express the overall analysis stage prior as a probability mass function: 
\[
p_i = \psi \text{Unif}[p_i, p_i] + (1 - \psi) \text{Beta}(\alpha_i, \beta_i),
\quad i = 1,2,
\]
where $\psi$ denotes the binary indicator variable for knowing exact values of $p_i$
beforehand. If $\psi = 1$, we are drawing from the uniform
distribution under the assumption of exact priors. Otherwise, $\psi = 0$ and we draw from the beta distribution to evaluate the analysis stage objective.

There is also an additional route we can use to showcase overlapping
behaviors between the Bayesian and frequentist paradigms.
Recall the sample size formula for assessing
differences in proportions in the frequentist setting,
\[
  n = \frac{(z_{1-\alpha/2} + z_{\beta})^2(p_1(1-p_1) + p_2(1-p_2))}{(p_1 - p_2)^2},
  \]
where $n = n_1 = n_2$. Simple rearragements and noting that
$-(z_{1-\alpha/2} + z_{\beta}) = z_{1-\beta} - z_{1-\alpha/2}$ lead us
to obtain
\begin{multline*}
\frac{\sqrt{n}(p_1 - p_2)}{\sqrt{p_1(1-p_1) + p_2(1-p_2)}} + z_{1-\alpha/2} = z_{1-\beta}
\implies
\text{Power} = 1 - \beta
= \\ \Phi\left(\frac{\sqrt{n}(p_1 - p_2)}{\sqrt{p_1(1-p_1) + p_2(1-p_2)}} + z_{1-\alpha/2}\right).
\end{multline*}
In an ideal situation, we could determine suitable
parameters for $\alpha_i$, $\beta_i$,
$i = 1, 2$ to use as our Beta priors that would enable
demonstration of convergence towards the frequentist setting.
However, a key relationship to recognize
is that the Beta distribution is a conjugate prior of the
Binomial distribution.
There is a subtle advantage offered given that the Bayesian credible
interval bands are based upon
posterior parameters of the Beta distribution and the frequentist
confidence interval bounds are based upon the Binomial distribution.
Because of the conjugate relationship held by the Beta and Binomial distributions,
we are essentially assigning priors to parameters
in the Bayesian setting that the Binomial density in the frequentist
setting is conditioned upon.
Using the fact
that the normal distribution can be used to approximate binomial distributions for large
sample sizes given that the Beta distribution is approximately normal when its
parameters $\alpha$ and $\beta$ are set to be equal and large. We manually
choose such values and apply them to our simulation study.

\subsubsection{Simulation Results}
\begin{figure}[t]
  \centering
    \includegraphics[width=0.75\textwidth]{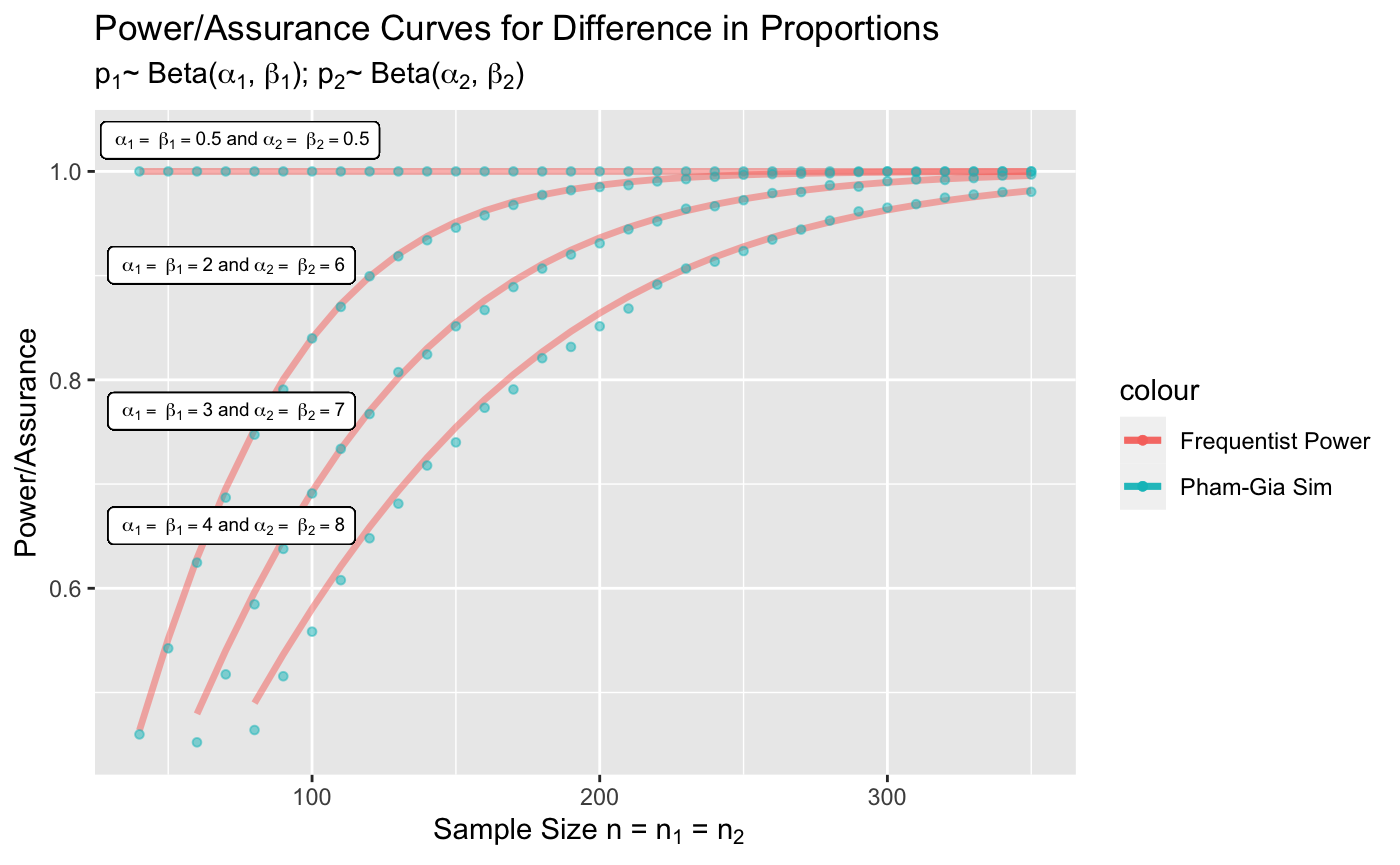}
    \caption{Overlay of simulated assurance results using posterior credible intervals and frequentist
    power results based on regular confidence intervals.}
  \label{fig:phamgiasim}
\end{figure}

Figure~\ref{fig:phamgiasim} displays the assurance curves
overlayed on top of the frequentist power curves. As mentioned in the previous section, we manually
set the parameters of the beta priors to be equal as doing so results in approximately
normal behavior. The horizontal line at the
top of the graph corresponds to flat priors for the beta distribution known as Haldane's
priors, in which the $\alpha$ and $\beta$ parameters are all set to 0.5.
Although the points do not align perfectly with the frequentist curves as we rely on an approximate relationship rather than
identifying prior assignments that allow direct ties to the frequentist case, our model still performs fairly well as the points and curves are still relatively close to one another.

\section{Discussion}\label{sec: conclusions}
This paper has attempted to present a simulation-based Bayesian design framework for sample size calculations using assurance for deciding in favor of a hypothesis (analysis objective). It is convenient to describe this framework in two stages: (i) the design stage generates data from a population modeled using design priors; and (ii) the analysis stage performs customary Bayesian inference using analysis priors. The frequentist setting emerges as a a special case of the Bayesian framework with highly informative design priors and completely uninformative analysis priors.

Our framework can be adapted and applied to a variety of clinical trial settings. Future directions of research and development can entail incorporating more complex analysis objectives into our framework. For example, the investigation of design and analysis priors in the use of Go/No Go settings \citep{Erik},
which refers to the point in time at which enough evidence is present to justify advancement to Phase 3 trials, will be relevant.
Whether the method of choice involves looking at lengths of posterior credible intervals \citep{joseph} or determining cutoffs that minimize the weighted sum of Bayesian average errors~\citep{reyes},
such conditions are all capable of being integrated as part of our analysis stage objective within
our two-stage paradigm.

\bibliographystyle{hapalike}
\bibliography{pow_samp}

\begin{thebibliography}{}

\bibitem[Adcock, 1997]{cj}
Adcock, C. (1997).
\newblock Sample size determination: A review.
\newblock {\em The Statistician}, 46(2).

\bibitem[Berger, 1985]{berger}
Berger, J.~O. (1985).
\newblock {\em Statistical Decision Theory and Bayesian Analysis}.
\newblock Springer New York, New York, NY.

\bibitem[Berry, 2006]{berry}
Berry, D.~A. (2006).
\newblock Bayesian clinical trials.
\newblock {\em Nature Reviews Drug Discovery}, 5(1).

\bibitem[Berry et~al., 2010]{berry2010}
Berry, S.~M., Carlin, B.~P., Lee, J.~J., and Müller, P. (2010).
\newblock {\em Bayesian Adaptive Methods for Clinical Trials}.
\newblock Chapman \& Hall/CRC Biostatistics Series, United Kingdom.

\bibitem[Cohen, 1988]{cohen}
Cohen, J. (1988).
\newblock {\em Statistical Power Analysis for the Behavioral Sciences (2nd
  ed.)}.
\newblock Lawrence Erlbaum Associates, Hillsdale, NJ.

\bibitem[Desu and Raghavarao, 1990]{desu}
Desu, M. and Raghavarao, D. (1990).
\newblock {\em Sample Size Methodology}.
\newblock Elsevier, Massachusetts.

\bibitem[Gelfand and Wang, 2002]{wang}
Gelfand, A.~E. and Wang, F. (2002).
\newblock A simulation based approach to bayesian sample size determination for
  performance under a given model and for separating models.
\newblock {\em Statistical Science}, 17(2).

\bibitem[Gelman et~al., 2013]{gelman}
Gelman, A., Carlin, J.~B., and Stern, H.~S. (2013).
\newblock {\em Bayesian Data Analysis (3rd ed.)}.
\newblock Chapman and Hall/CRC, United Kingdom.

\bibitem[Ibrahim et~al., 2012]{Ibrahim}
Ibrahim, J.~G., Chen, M.-H., Xia, A., and Liu, T. (2012).
\newblock Bayesian meta-experimental design: Evaluating cardiovascular risk in
  new antidiabetic therapies to treat type 2 diabetes.
\newblock {\em Biometrics}, 68(2).

\bibitem[Joseph et~al., 1997]{joseph}
Joseph, L., Berger, R.~D., and Belisle, P. (1997).
\newblock Bayesian and mixed bayesian/likelihood criteria for sample size
  determination.
\newblock {\em Statistics in Medicine}, 16(7).

\bibitem[Kraemer and Thiemann, 1987]{thiemann}
Kraemer, H.~C. and Thiemann, S. (1987).
\newblock {\em How Many Subjects? Statistical Power Analysis in Research}.
\newblock Sage Publications, Newbury Park.

\bibitem[Lee and Chu, 2012]{lee}
Lee, J. and Chu, C.~T. (2012).
\newblock Bayesian clinical trials in action.
\newblock {\em Statistics in Medicine}, 31(25).

\bibitem[Lee and Zelen, 2000]{lee_zelen}
Lee, S.~J. and Zelen, M. (2000).
\newblock Clinical trials and sample size considerations: Another perspective.
\newblock {\em Statistical Science}, 15(2).

\bibitem[Lindley, 1997]{lindley}
Lindley, D.~V. (1997).
\newblock The choice of sample size.
\newblock {\em The Statistician}, 46(2).

\bibitem[Liu and Liang, 1997]{liu}
Liu, G. and Liang, K.~Y. (1997).
\newblock Sample size calculations for studies with correlated observations.
\newblock {\em Biometrics}, 53(3).

\bibitem[Muller et~al., 1992]{muller}
Muller, K.~E., Lavange, L.~M., Ramey, S.~L., and Ramey, C.~T. (1992).
\newblock Power calculations for general linear multivariate models including
  repeated measures applications.
\newblock {\em Journal of the American Statistical Association}, 87(420).

\bibitem[Müller and Parmigiani, 1995]{parm}
Müller, P. and Parmigiani, G. (1995).
\newblock Optimal design via curve fitting of monte carlo experiments.
\newblock {\em Journal of the American Statistical Association}, 90(432).

\bibitem[O'Hagan and Stevens, 2001]{ohagan}
O'Hagan, A. and Stevens, J.~W. (2001).
\newblock Bayesian assessment of sample size for clinical trials of
  cost-effectiveness.
\newblock {\em Medical Decision Making}, 21(3).

\bibitem[Parmigiani, 2002]{parmigiani}
Parmigiani, G. (2002).
\newblock {\em Modeling in Medical Decision Making: A Bayesian Approach}.
\newblock Wiley, Hoboken, NJ.

\bibitem[Pham-Gia, 1997]{pham}
Pham-Gia, T. (1997).
\newblock On bayesian analysis, bayesian decision theory and the sample size
  problem.
\newblock {\em The Statistician}, 46(2).

\bibitem[Pulkstenis et~al., 2017]{Erik}
Pulkstenis, E., Patra, K., and Zhang, J. (2017).
\newblock A bayesian paradigm for decision-making in proof-of-concept trials.
\newblock {\em Journal of Biopharmaceutical Statistics}, 27(3).

\bibitem[Rahme et~al., 2000]{rahme}
Rahme, E., Joseph, L., and Gyorkos, T.~W. (2000).
\newblock Bayesian sample size determination for estimating binomial parameters
  from data subject to misclassification.
\newblock {\em Journal of Royal Statistical Society}, 49(1).

\bibitem[Raiffa and Schlaifer, 1961]{raiffa}
Raiffa, H. and Schlaifer, R. (1961).
\newblock {\em Applied Statistical Decision Theory}.
\newblock Harvard University Graduate School of Business Administration
  (Division of Research), Massachusetts.

\bibitem[Reyes and Ghosh, 2013]{reyes}
Reyes, E.~M. and Ghosh, S.~K. (2013).
\newblock Bayesian average error based approach to sample size calculations for
  hypothesis testing.
\newblock {\em Biopharm}, 23(3).

\bibitem[Self and Mauritsen, 1988]{self}
Self, S.~G. and Mauritsen, R.~H. (1988).
\newblock Power/sample size calculations for generalized linear models.
\newblock {\em Biometrics}, 44(1).

\bibitem[Self et~al., 1992]{ohara}
Self, S.~G., Mauritsen, R.~H., and O'Hara, J. (1992).
\newblock Power calculations for likelihood ratio tests in generalized linear
  models.
\newblock {\em Biometrics}, 48(1).

\bibitem[Spiegelhalter et~al., 1993]{spiegelhalter}
Spiegelhalter, D.~J., Freedman, L.~S., and Parmar, M.~K. (1993).
\newblock Applying bayesian ideas in drug development and clinical trials.
\newblock {\em Statistics in Medicine}, 12(15).

\end{thebibliography}

  \end{document}